\documentclass[a4paper,12pt]{article}
\usepackage{emptypage} 
\usepackage{a4wide}

\usepackage[pagestyles]{titlesec}

\usepackage{amsmath}

\usepackage{accents}

\newpagestyle{main}{
\headrule
\sethead[\thepage][\chaptertitle][] 
{}{\sectiontitle}{\thepage} 
}

\usepackage{latexsym}
\usepackage{amssymb}
\usepackage{amsthm}

\usepackage{graphicx}

\usepackage{amstext}
\usepackage{bbm}

\usepackage{xcolor}
\usepackage{hyperref}

\usepackage[capitalize,nameinlink]{cleveref}
\usepackage{autonum}

\usepackage[]{currvita}

\usepackage[math]{cellspace}
\makeatletter
\def\underbracex#1#2{\mathop{\vtop{\m@th\ialign{##\crcr
   $\hfil\displaystyle{#2}\hfil$\crcr
   \noalign{\kern3\p@\nointerlineskip}%
   #1\crcr\noalign{\kern3\p@}}}}\limits}

\def\upbracefilla{$\m@th \setbox\z@\hbox{$\braceld$}%
  \bracelu\leaders\vrule \@height\ht\z@ \@depth\z@\hfill 
\kern\p@\vrule \@width\p@\kern\p@\vrule \@width\p@\kern\p@\vrule \@width\p@
$}

\def\upbracefillb{$\m@th \setbox\z@\hbox{$\braceld$}%
\vrule \@width\p@\kern\p@\vrule \@width\p@\kern\p@\vrule \@width\p@\kern\p@
 \leaders\vrule \@height\ht\z@ \@depth\z@\hfill\bracerd
  \braceld\leaders\vrule \@height\ht\z@ \@depth\z@\hfill
\kern\p@\vrule \@width\p@\kern\p@\vrule \@width\p@\kern\p@\vrule \@width\p@
$}

\def\upbracefillc{$\m@th \setbox\z@\hbox{$\braceld$}%
\vrule \@width\p@\kern\p@\vrule \@width\p@\kern\p@\vrule \@width\p@\kern\p@
\leaders\vrule \@height\ht\z@ \@depth\z@\hfill
\kern\p@\vrule \@width\p@\kern\p@\vrule \@width\p@\kern\p@\vrule \@width\p@
$}

\def\upbracefill{$\m@th \setbox\z@\hbox{$\braceld$}%
\vrule \@width\p@\kern\p@\vrule \@width\p@\kern\p@\vrule \@width\p@\kern\p@
 \leaders\vrule \@height\ht\z@ \@depth\z@\hfill\braceru$}

\def\upbracefillL{$\m@th \setbox\z@\hbox{$\braceld$}%
  \bracelu\leaders\vrule \@height\ht\z@ \@depth\z@\hfill 
\kern\p@\vrule \@width\p@\kern\p@\vrule \@width\p@\kern\p@\vrule \@width\p@
$}

\def\upbracefillR{$\m@th \setbox\z@\hbox{$\braceld$}%
\vrule \@width\p@\kern\p@\vrule \@width\p@\kern\p@\vrule \@width\p@\kern\p@
 \leaders\vrule \@height\ht\z@ \@depth\z@\hfill\bracerd
  \braceld\leaders\vrule \@height\ht\z@ \@depth\z@\hfill
 \leaders\vrule \@height\ht\z@ \@depth\z@\hfill\braceru$}

\def\upbracefillLR{$\m@th \setbox\z@\hbox{$\braceld$}%
  \bracelu\leaders\vrule \@height\ht\z@ \@depth\z@\hfill 
\leaders\vrule \@height\ht\z@ \@depth\z@\hfill\bracerd
  \braceld\leaders\vrule \@height\ht\z@ \@depth\z@\hfill
 \leaders\vrule \@height\ht\z@ \@depth\z@\hfill\braceru
$}
\makeatother

\def\R{\mathbb R}

\def\N{\mathbb N}

\newcommand{\p}{\psi}
\renewcommand{\t}{\theta}

\newcommand{\ti}{\tilde }

\newcommand{\tixi}{\ti \xi}
\newcommand{\tiu}{\ti \u}

\newcommand{\hxi}{\hat \xi}
\newcommand{\hu}{\hat \u}

\def\u{u}
\def\g{\gamma}
\def\gz{\gamma(\u_0)}

\def\a{\alpha}

\renewcommand{\v}{v}
\newcommand{\w}{w}

\newcommand{\tztt}{\theta_{0}(\xi(t),\u(t),x)}
\newcommand{\pztt}{\psi_{0}(\xi(t),\u(t),x)}

\newcommand{\tzt}{\theta_{0}(\xi,\u,x)}
\newcommand{\pzt}{\p_{0}(\xi,\u,x)}

\newcommand{\pa}{\partial_}

\newcommand{\tp}{\l(\begin{matrix}
\t\\
\p\\
\end{matrix}\r)}

\newcommand{\tpP}{\l(\begin{matrix}
\t'\\
\p'\\
\end{matrix}\r)}

\newcommand{\dt}{\pa t}
\newcommand{\ddt}{\fr d{dt}}

\newcommand{\ds}{\fr d {ds}}

\newcommand{\dxi}{\pa \xi}

\newcommand{\du}{\pa \u}

\newcommand{\dx}{\pa x}
\newcommand{\dZ}{\pa Z}
\newcommand{\deps}{\pa \eps}

\renewcommand{\d}{\dot}

\def \F{F(\eps,x)}

\newcommand{\OM}[2]{\Omega \l({{#1}},{{#2}}\r)}
\newcommand{\OT}[2]{\stackrel{{\smash{\scriptscriptstyle{#2}}}}{{#1}}}
\newcommand{\Ltwo}[2]{\l\langle {#1},{#2} \r\rangle_{L^2(\R)\oplus L^2(\R)}}

\newcommand{\tnv}[3]{\l|{#1}\r|_{L^\infty({[{#2},{#3}]},H^1(\R))}}
\newcommand{\tnvinf}[3]{\l|{#1}\r|_{L^\infty(\R)L^\infty({[{#2},{#3}]})}}

\newcommand{\tnw}[3]{\l|{#1}\r|_{L^\infty({[{#2},{#3}]},L^2(\R))}}
\newcommand{\nw}[1]{\l|{#1}\r|_{L^2(\R)}}

\newcommand{\nltwo}[1]{\l|{#1}\r|_{L^2(\R)}}
\newcommand{\nlinf}[1]{\l|{#1}\r|_{L^\infty(\R)}}

\newcommand{\nhone}[1]{\l|{#1}\r|_{H^1(\R)}}

\newcommand{\CR}{\color{black}}
\newcommand{\fr}{\frac}

\def\be#1\ee{\begin{align}#1\end{align}}
\def\ben#1\een{\begin{align+}#1\end{align+}}
\def\ba#1\ea{\begin{aligned}[t]#1\end{aligned}}
\def\bs#1\es{\begin{split}#1\end{split}}

\renewcommand{\l}{\left}
\renewcommand{\r}{\right}
\def\?{????}

\newcommand{\bma}{\begin{pmatrix}}
\newcommand{\ema}{\end{pmatrix}}

\newcommand{\bmat}{\begin{bmatrix}}
\newcommand{\emat}{\end{bmatrix}}

\newcommand{\bca}{\begin{cases}}
\newcommand{\eca}{\end{cases}}

\def\J{\mathbb J}
\newcommand{\re}{\eqref}
\newcommand{\nn}{\nonumber}

\newcommand{\la}{\label}
\newcommand{\beq}{\begin{equation}}
\newcommand{\eeq}{\end{equation}}
\newcommand{\eps}{\varepsilon}

\renewcommand{\qed}{\protect~\protect\hfill $\Box$}

\begin{document}


\newtheoremstyle{plainNEW}
  {}
  {}
  {\itshape}
  {}
  {\boldmath\bfseries}
  {.}
  { }
  {\thmname{#1}\thmnumber{ #2}\thmnote{ (#3)}}%


\theoremstyle{plainNEW}

\newtheorem{theorem}{Theorem}[section]

\newtheorem{definition}[theorem]{Definition}
\newtheorem{deflem}[theorem]{Definition and Lemma}
\newtheorem{lemma}[theorem]{Lemma}
\newtheorem{corollary}[theorem]{Corollary}
\newcommand{\bde}{\begin{definition}}
\newcommand{\ede}{\end{definition}}
\newcommand{\ble}{\begin{lemma}}
\newcommand{\ele}{\end{lemma}}
\newcommand{\bre}{\begin{remark}}
\newcommand{\ere}{\end{remark}}
\newcommand{\bco}{\begin{corollary}}
\newcommand{\eco}{\end{corollary}}
\newcommand{\bpro}{\begin{proposition}}
\newcommand{\epro}{\end{proposition}}

\newtheorem{assumption}{Assumptions}[theorem]
\newcommand{\bas}{\begin{assumption}}
\newcommand{\eas}{\end{assumption}}

\theoremstyle{definition}

\newtheorem{example}[theorem]{Example}
\newtheorem{remark}[theorem]{Remark}
\newtheorem{remarks}[theorem]{Remarks}
\newtheorem*{proofNEW}{Proof}
\newcommand{\bpr}{\begin{proofNEW}}
\newcommand{\epr}{\qed
\bigskip
\end{proofNEW}}

\theoremstyle{plainNEW}

\newcommand{\bth}{\begin{theorem}}
\renewcommand{\eth}{\end{theorem}}
\newtheorem{proposition}[theorem]{Proposition}

\pagenumbering{gobble}
 
\newpage
\title{
Solitons in the Presence of a Small, Slowly Varying Electric Field
}

\date{}

\author{{\sc Timur Mashkin}\\[2ex]
         Mathematisches Institut, Universit\"at K\"oln, \\
         Weyertal 86-90, D\,-\,50931 K\"oln, Germany \\
         e-mail: tmashkin@math.uni-koeln.de}

\maketitle

\begin{abstract}\noindent
We consider the perturbed sine-Gordon equation $\theta_{tt}-\theta_{xx}+\sin \theta= 
 \eps^2 f(\eps x)$, where the external perturbation $\eps^2 f(\eps x)$  corresponds to a small, slowly varying electric field.
We show that the initial value problem with an appropriate initial state 
close enough to the solitary manifold has a unique solution, which follows up to time $1/{\eps^{}}$ and errors of order $\eps^{ 3/ 4}$ a trajectory on the solitary manifold. 
The trajectory on the solitary manifold is described by ODEs, which agree 
up to errors of order $\eps^3$
with 
Hamilton equations for the 
restricted to the solitary manifold 
sine-Gordon
Hamiltonian.
\smallskip

\end{abstract}

\pagenumbering{arabic}

\section{Introduction}
\noindent 
The perturbed sine-Gordon equation 
\beq\la{SGE}
\t_{tt}-\t_{xx}+\sin\t= F(\eps,x),~~~~t,x\in\R,~~~~\eps\ll 1,
\eeq
is a Hamiltonian evolution equation with Hamiltonian given by
\be\la{intro Hamiltonian}
H^\eps(\t,\p)=\fr 1 2\int \p^2+\t_x^2+2(1-\cos\t)-2F(\eps,x)\t \,dx\, 
\ee
and the symplectic form given by
\be\la{symplecticform introduction}
\Omega\l(\tpP,\tp \r) = \l\langle \tpP,\J\tp \r\rangle_{L^2(\R)\oplus L^2(\R)}=\int_\R \p'(x)\t(x)-\t'(x)\p(x)\,dx,
\ee
where 
$$\J=\l(\begin{matrix}
0&-1\\
1&0\\
\end{matrix}\r).$$ 
In first order formulation \re{SGE} can be written as a system:
\be\la{SGE1 first order introduction}
\partial_t\bma
\t\\
\p
\ema=\l(\begin{matrix}
\p\\
\t_{xx}-\sin\t+F(\eps, x)\\
\end{matrix}\r).
\ee
The unperturbed sine-Gordon equation ($F=0$),
admits soliton solutions 
$
\bma 
\t_0(\xi(t),\u(t),x)\\
\p_0(\xi(t),\u(t),x)
\ema
$,
where
$$
\dot\xi=\u\,, ~~ \dot \u=0\,,~~~~(\xi(0),\u(0))=(a,v)\in\R\times(-1,1).
$$
Here the functions $(\t_0,\p_0)$ are defined by
\be\la{solitonsolution}
{}&\bma
\t_0(\xi,\u,x)\\
\p_0(\xi,\u,x)
\ema:=\bma
\t_K(\g(u)(x-\xi))\\
-\u\g(u)\t_K'(\g(u)(x-\xi))\\
\ema\,,~\u\in(-1,1),~~\xi,x\in\R,
\ee
where 
$$
\g(u)=\frac 1 {\sqrt{1-u^2}},~~~~\t_K(x) =4\arctan(e^x),
$$
and $\t_K$ satisfies $\t_K''(x)=\sin\t_K(x)$ with boundary conditions $\t_K(x) \to \bma 2\pi\\0  \ema$ as $x\to \pm \infty$.
The states $\l(\begin{matrix}
\t_0(a,v,\cdot)\\
\p_0(a,v,\cdot)\\
\end{matrix}\r)$ form the  two-dimensional
solitary manifold  
\be
\la{intro classical manifold}
{\cal S}_0:=\l\{ \l(\begin{matrix}
\t_0(a,v,\cdot)\\
\p_0(a,v,\cdot)\\
\end{matrix}\r)~:~v\in(-1,1),~a\in\R
\r\}.
\ee
The sine-Gordon equation arises in various physical phenomena such as dynamics of long Josephson junctions \cite{0953-8984-7-2-013,RevModPhys.61.763}, dislocations in crystals 
\cite{FrenKont}, waves in ferromagnetic materials \cite{0022-3719-11-1-007}, etc. T. H. R. Skyrme \cite{Skyrme237} proposed the equation to model elementary particles. Dynamics of solitons under constant electric field were examined numerically, for instance, in \cite{doi:10.1143/JPSJ.46.1594}.
In this paper, we investigate solitons in the presence of a time independent, small, slowly varying electric field of the type $F(\eps,x)=\eps^2 f(\eps x)$, which is physically relevant.
Our main result is the following theorem.
\bth\la{maintheorem epstwof}
Let $f\in H^3(\R)$, $0<U<1$, $(\xi_s,\u_s)\in \R \times (-U,U)$ and
$m= \int  (\t_K'(Z))^2\,dZ$. 
We consider the Cauchy problem
\be\la{SGE1 initialdataprob introduction}
\partial_t\bma
\t\\
\p
\ema=\l(\begin{matrix}
\p\\
\t_{xx}-\sin\t+\eps^2 f(\eps x)\\
\end{matrix}\r)~~
\bma
\t(0,x)\\
\p(0,x)
\ema =
\bma
\t_0(\xi_s,\u_s,x)\\
\p_0(\xi_s,\u_s,x)
\ema+
\bma
\v(0,x)\\
\w(0,x), 
\ema
\ee
such that the following assumptions are satisfied:
\begin{itemize}
\item[(a)] $\eps$ is sufficiently small. 
\item[(b)] ${\cal N}(\t(0,x),\p(0,x),\xi_s,\u_s)=0\,$, 
where 
$
{\cal N} =({\cal N}_1,{\cal N}_2) : L^\infty(\R) \times L^2(\R)\times \Sigma(2,U)\to\R^2  
$ is given by 
\be\la{orthogonalitycondIntro}
{\cal N}(\t,\p,\xi,\u)
:=
\bma
\OM{\bma
\dxi\t_0(\xi,\u,\cdot)\\
\dxi\p_0(\xi,\u,\cdot)
\ema}{\bma
\t(\cdot)-\t_0(\xi,\u,\cdot)\\
\p(\cdot)-\p_0(\xi,\u,\cdot)\\
\ema}\\
\OM{\bma
\du\t_0(\xi,\u,\cdot)\\
\du\p_0(\xi,\u,\cdot)\\
\ema}{\bma
\t(\cdot)-\t_0(\xi,\u,\cdot)\\
\p(\cdot)-\p_0(\xi,\u,\cdot)\\
\ema}
\ema\, 
\ee
with $V(l):= \fr{1-U} l$, $\Sigma(l,U):=\Big\{(\xi,\u)\in \R\times (-1,1): 
u\in (-U- V(l),U+ V(l) ) \Big\}$, 
%
and the symplectic form $\Omega$ given by \re{symplecticform introduction}.

\item[(c)] $\nhone{v(0,\cdot)}^2+\nltwo{w(0,\cdot)}^2\le \eps^{\fr {11} 4}$, where $(\v(0,\cdot),\w(0,\cdot))$ is given by \re{SGE1 initialdataprob introduction}.

\end{itemize}
The Cauchy problem 
has a unique solution on the time interval
\be
0\le t\le T, ~\text{where}~ 
T:=T(\eps):=
\fr {1} {\eps^{}}.
\ee
The solution may be written in the form
\be 
\bma
\t(t,x)\\ 
\p(t,x)
\ema=\bma 
\t_0( \bar\xi(t), \bar\u(t),x)\\ 
\p_0( \bar\xi(t), \bar\u(t),x)
\ema+ 
\bma
\v(t,x)\\
\w(t,x)
\ema,
\ee
where $\v, \w$
have regularity
\be
&(v(t), w(t)) \in C^1([0,T ] , H^1(\R) \oplus L^2(\R)),
\ee
$\bar\xi,\bar\u$ solve the following system of equations 
\be\la{exactODE epstwof1} 
 \d{ \bar\xi}( t) =   \bar\u( t) \,,~~~~
  \d{\bar\u}( t) =  -\eps^2\fr { f(\eps\bar\xi( t))} {[\g(\bar\u( t))]^3 m}\int \t_K ' (Z) \,dZ  \,,
\ee
with initial data 
$
\bar\xi(0)=\xi_s,~\bar\u(0)=u_s,
$
and there exists a positive constant $c$ such that
\be\la{bound on vw intro}
\tnv{v}{0}{T}^2+\tnw{w}{0}{T}^2\le c \eps^{\fr {3}2}\,.
\ee
The constant $c$ depends on $f$. The parameters $\bar\xi,\bar\u$ describe
a fixed nontrivial perturbation of the uniform linear motion
as $\eps \to 0$ if condition $f(0)\not=0$ is satisfied.
\eth
This result yields a fairly accurate description of the solution $(\t,\p)$ to the Cauchy problem, since we are able to control
the dynamics of the transversal component $(v(t,\cdot),w(t,\cdot))$ by the upper bound on its norm \re{bound on vw intro} and the dynamics on the solitary manifold ${\cal S}_0$ by the 
ODEs \re{exactODE epstwof1}.
The time scale is nontrivial and the result provides a nontrivial dynamics on the solitary manifold ${\cal S}_0$ 
as $\eps \to 0$ if $f(0)\not=0$. 

%
Let us mention some related works. Orbital stability of soliton solutions 
under perturbations of the initial data
has been proven 
for the unperturbed sine-Gordon equation 
(see \cite{MR678151}, \cite[Section 4]{Stuart3}).
D. M. Stuart \cite{Stuart2} considered also the perturbed sine-Gordon equation
\be
{}&\t_{TT}-\t_{XX}+\sin\t+\eps g=0,\nn
\ee
where the perturbation $g=g(\t)$ is a smooth function such that $g_0(Z)=g(\t_K(Z))\in L^2(dZ)$ and $\eps\ll 1$. He proved  that there exists $T_* = {\cal O} \l( \fr 1 \eps \r) $ such that the corresponding initial value problem with initial data
\be
{}&\t(0,X)=\t_K(Z(0))+\eps \ti\t(0,X),~~~~
\t_T(0,X)=\fr {-u(0)}{\sqrt{1-u(0)^2}}\t_K'(Z(0))+\eps \ti\t_T(0,X),\nn\\
{}& (\ti\t(0,X),\ti\t_T(0,X))\in H^1\oplus L^2,\nn
\ee
has a unique solution of the form 
\be
\t(T,X)=\t_K(Z)+\eps\ti\t(T,X),~~~Z=\fr{X-\int^T u - C(T)}{\sqrt{1-u^2}},\nn
\ee
where $\ti\t\in C([0,T_*],H^1),~\t_T\in C([0,T_*],L^2)$ and
\be
{}&C(T)=C_0(\eps T)+\eps \ti C,~~
{}&u(T)=u_0(\eps T)+\eps \ti\u(T) \l( \Rightarrow~ p= \fr {u} {\sqrt{1-u^2}}= p_0(\eps T)+\eps \ti p(T)  \r).\nn
\ee
Here $\ti p,~\ti u,\ti C, \fr {d\ti u}{d T}, \fr {d\ti C}{d T}, |\ti\t |_{H^1(\R)}$ are bounded independent of $\eps$. The functions $u_0,C_0$ are solutions of 
certain explicitly given modulation equations.
This result is also valid for perturbations $g$ of the  form
$$
g=g(\eps T, \eps X, \t),
$$
if among others the following assumption is satisfied:
There exists a time interval $
\l[ 0, \fr{t_+} \eps\r]$ and a constant $A$ such that for all $T\in \l[ 0, \fr{t_+} \eps\r] $:
\be
\l(\int  g(\eps T, \eps X, \t_K(Z))^2 \,dZ\r)^\fr 12 \le A ,~~~dZ=\g(u) dX,~~~\g(u)=1/\sqrt {1-u^2}, \la{conditionStuart introduction}
\ee
where $t_+$ and $A$ are independent of $\eps$
(see \cite[p. 442]{Stuart2}). The proof is based on an orthogonal
decomposition of the solution into an oscillatory part and a one-dimensional
"zero-mode" term.

In \cite{MashkinDissertation} we studied equation \re{SGE1 first order introduction} with the perturbation $F(\eps,x)= \eps f(\eps x)$, where $f\in H^1(\R)$.   
We proved 
that the corresponding Cauchy problem with initial data $\eps$-close to the 
solitary manifold  
$
{\cal S}_0  
$
has a unique solution which follows up to time $1/\eps^{\fr 1 4}$ and errors of order $\eps $ a trajectory
on 
$
{\cal S}_0  
$. The trajectory on $
{\cal S}_0  
$ is described by parameters which satisfy
ODEs for uniform linear motion.  
Notice that the perturbation $F(\eps,x)= \eps f(\eps x)$ in \cite{MashkinDissertation} is not comparable to the perturbations considered in \cite{Stuart2}, since $F(\eps,x)=\eps f(\eps x)$ does not depend on time and the condition \re{conditionStuart introduction} is not satisfied due to
  $$\nltwo{f(\eps \cdot)}=\eps^{-\fr1 2} \nltwo{f( \cdot)}.
	$$
  
The main result of 
the present
paper 
yields 
richer dynamics on the solitary manifold than the mentioned result of \cite{MashkinDissertation} in the following sense: 
In the present paper the parameters $\bar\xi,\bar\u$
satisfy ODEs which describe nontrivial perturbation of the straight line unperturbed
dynamics if $f(0)\not=0$, whereas in \cite{MashkinDissertation} the corresponding ODEs describe merely straight line unperturbed
dynamics.
So in 
this paper
the influence of the electric field  $F(\eps,x)$ is identifiable in the dynamics on the solitary manifold in contrast to the mentioned result of \cite{MashkinDissertation}. 
The richer dynamics can be captured, among others,
due to the fact that
we are able 
to reach the dynamically relevant time
frame $\eps^{-1}$ (see proof of \cref{le: prepairing main proof epstwof}).

%

In \cite{Mashkin} we considered the sine-Gordon equation with perturbations
$$F: (-1,1)\to H^{1,1}(\R), ~\eps \mapsto F(\eps,\cdot),$$
of class
$
C^{n}((-1,1),H^{1,1}(\R))
$ 
whose first $k$ derivatives vanish at 0, i.e., 
$\deps^l F(0,\cdot)=0~~ \text{for} ~~0\le l\le k$, where $k+1 \le n$ and $n\ge 1$.
We constructed there, by successive deformation of the classical solitary manifold ${\cal S}_0$, 
a virtual solitary manifold ${\cal S}_n^\eps$, 
which was obtained in $n$ iteration steps.
The virtual solitary manifold ${\cal S}_n^\eps$
is defined 
by an implicitly given function
$(\t_n^\eps(a,v,x),
\p_n^\eps(a,v,x)
)$ 
analogous to \re{intro classical manifold}
and it is adjusted to the perturbation $F$. 
The result of \cite{Mashkin}, which was obtained by using
the concept of virtual solitons and Lyapunov functional methods, is as follows:
For $\xi_s\in\R$, $\eps\ll 1$
the initial value problem
\be
\partial_t \bma
\t \\
\p 
\ema{}&=\l(\begin{matrix}
\p \\
\dx^2\t -\sin\t +\F\\
\end{matrix}\r),~
\bma
\t(0,x)\\
\p(0,x)
\ema{}&=
\bma
\t^\eps_n(\xi_s,\u_s,x)\\
\p^\eps_n(\xi_s,\u_s,x)
\ema+
\bma
\v(0,x)\\
\w(0,x) 
\ema,
\ee
with appropriate initial data that is $\eps^n$-close to ${\cal S}_n^\eps$, i.e., 
$
\nhone{v(0,\cdot)}^2+\nw{w(0,\cdot)}^2\le \eps^{2n},
$ 
with initial velocity that satisfies the smallness assumption
$
|u_s|\le \ti C \eps^{\fr{k+1}2} 
$,
has a unique solution $(\t,\p)$  
which may be written up to time
$  1/ (\ti C{\eps} ^{\fr{k+1}2})$ in the form
\be
\bma
\t(t,x)\\ 
\p(t,x)
\ema=\bma 
\t_n^\eps(\bar\xi(t),\bar\u(t),x)\\ 
\p_n^\eps(\bar\xi(t),\bar\u(t),x)
\ema+ 
\bma
\v(t,x)\\
\w(t,x)
\ema.\nn
\ee
The solution remains $\eps^n$-close to ${\cal S}_n^\eps$, i.e.,
$
\nhone{v(t,\cdot)}^2+\nltwo{w(t,\cdot)}^2\le \ti C \eps^{2n} ,
$
and the dynamics on ${\cal S}_n^\eps$ is described precisely by the parameters $(\bar\xi(t),\bar\u(t))$ which satisfy exactly the ODEs
\be\la{ODE introduction}
 \d{\bar\xi}( t) =  \bar\u(t)  \,,~~~~
 \d{\bar\u}( t) = \lambda_{n}^\eps\l(\bar\xi(t), \bar\u(t)\r), 
\ee
with initial data 
$
\bar\xi(0)=\xi_s,~\bar\u(0)=\u_s
$, where the function $\lambda_n^\eps$ is given implicitly.
The 
parameters $\bar\xi,\bar\u$ describe
a fixed nontrivial perturbation of the uniform linear motion
as $\eps \to 0$ if the perturbation $F$  satisfies a specific condition.

The result of \cite{Mashkin} 
can be applied to  
perturbations of type $F(\eps,x)=\eps^2f(\eps x)$ with appropriate $f$ by choosing $n=1$ and $k=0$. This yields a stability statement with a shorter time scale than in the main result of the present paper and with an additional smallness assumption on the initial velocity. 
In this paper we make use of the fact that the perturbation $F(\eps,x)=\eps^2f(\eps x)$ is slowly varying (see for instance proof of \cref{le: prepairing main proof epstwof}), which is one of the reasons for the differences in the statements of both results.    

There exist also many results on
stability of solitons for several other equations. 
For instance,
J. Holmer and M. Zworski considered in \cite{HoZwSolitonint} the Gross-Pitaevskii equation 
$i\partial_t u + \tfrac{1}{2}\partial_x^2 u - V ( x ) u +u|u|^2 = 0,$
with a slowly varying smooth potential $V(x)=W(\eps x)$
and proved that up to time $\fr{\log(1/\eps)}\eps$ and errors of size $\eps^2$ in $H^1$, the solution is a soliton evolving according to the classical dynamics of a natural effective Hamiltonian. 

Further long (but finite)-time results for different equations with external potentials  can be found in \cite{MR2094474,MR2232367,MR2342704,MR2855072}.
For 
results on orbital stability and long time soliton asymptotics 
see
for example 
\cite{MR820338,MR0428914,MR0386438,MR2920823,MR1071238,MR1221351,ImaykinKomechVainberg,MR3630087,MR3461359}. 

The main result of this paper is based on \cite[Part II]{Mashkin}, where many of the computations are presented in greater detail.

Let us comment on our techniques.
The local solution of \re{SGE1 initialdataprob introduction} exists due to the contraction mapping theorem. By the following approach we derive some estimates which imply that the local solution is continuable and that the bound stated in \cref{maintheorem epstwof} is satisfied.\\
We decompose the solution of \re{SGE1 initialdataprob introduction} into a point on the virtual solitary manifold ${\cal S}_0$ and a transversal component which is symplectic orthogonal to the tangent space of ${\cal S}_0$ at the corresponding point:
\be\la{intro decompositionofthesolution}
\bma
\t(t,x)\\ 
\p(t,x)
\ema=\bma 
\t_0( \xi(t), \u(t),x)\\ 
\p_0( \xi(t), \u(t),x)
\ema+ 
\bma
\v(t,x)\\
\w(t,x)
\ema.
\ee
This symplectic decomposition is possible in a small uniform distance to the solitary manifold due to the implicit function theorem.
The energy
\be\la{energy introduction}
{}& H(\t,\p)=\fr 1 2\int \p^2+\t_x^2+2(1-\cos\t) \,dx
\ee
and the momentum 
\be\la{momentum introduction}
{}&\Pi (\t ,\p )
 = \int\p\t_x\,dx 
\ee
are conserved quantities of the unperturbed sine-Gordon equation. We make use of this fact and achieve control over the transversal component $(v,w)$ of the solution by introducing an almost conserved
Lyapunov function, given by  
\be 
 L =\int \fr{\w^2 } 2 +\fr{\dx\v^2} 2 +\fr{\cos(\t_0(\xi,\u,\cdot)) \v^2 }2+\u\w \dx\v \,dx,\nn
\ee
where $(v,w)$ and $(\xi,\u)$ are such as in \re{intro decompositionofthesolution}.
$L $ is the quadratic part of 
$$
H(\t_0+v,\p_0+w)+u\Pi(\t_0+v,\p_0+w) - \Big (H(\t_0,\p_0)+u\Pi(\t_0,\p_0) \Big),
$$
where the linear part vanishes due to symplectic orthogonality in decomposition \re{intro decompositionofthesolution}. 
The Lyapunov function is bounded from below in terms of $\nhone{v(t,\cdot)}^2+\nltwo{w(t,\cdot)}^2$, which is a consequence of the symplectic orthogonality in the decomposition and of spectral properties of the operator $-\dZ^2+\cos\t_K(Z)$. 

To obtain the ODEs \re{exactODE epstwof1} we follow the idea of \cite{HoZwSolitonint}. Namely, we compute the flow of the Hamiltonian $H^\eps$ restricted to the solitary manifold ${\cal S}_0$ and discard all terms of order $\eps^3$ or higher.
The parameters $(\xi ,\u )$ from \re{intro decompositionofthesolution} satisfy the ODEs \re{exactODE epstwof1} up to errors of order $\eps^\fr{11}4$.
This will be used in order to control from above the Lyapunov function and consequently also the norm of the transversal component $(v,w)$. 
Using Gronwall's lemma we pass from the approximate equations for the parameters $(\xi,\u)$
to the exact ODEs \re{exactODE epstwof1}.
It suffices to consider the flow of the restricted Hamiltonian $H^\eps$ without terms of order $\eps^3$ (or higher), since the fact that $(\xi,\u)$
satisfies ODEs \re{exactODE epstwof1} up to mentioned orders enables us to carry out all computations leading to the main result. 

The ODEs \re{exactODE epstwof1} can be rescaled in time by introducing $s=\eps t$,
$
\hxi(s):= \bar\xi(s/\eps^{   })
$,
$
\hu(s):= \fr 1 {\eps^{  }} {\bar u(s/\eps^{  })}
$. 
The 
corresponding transformed ODEs have the form 
\be
\ds \hxi(s) =   \hu(s) \,,~~~~~~
\ds \hu(s) =  - \fr {f(\eps\hxi(s))} {[\g(\eps\hu(s))]^3 m} \int \t_K ' (Z) \,dZ   
\ee
and converge to ODEs that describe a fixed nontrivial perturbation of the uniform linear motion as $\eps \to 0$ if $f(0)\not=0$. 

The paper is organized as follows.  
We prove in \cref{Symplectic Orthogonal Decomposition epsf} that in a uniform distance to the solitary manifold the decomposition into symplectically orthogonal components is possible. The existence of a local solution $(\t,\p)$ with initial state close to the solitary manifold is established in \cref{local solution epstwof}. 
In \cref{Modulation Equations for ODE analysis epstwof} we show that the parameters $(\xi,\u)$ 
from the symplectic decomposition \re{intro decompositionofthesolution} 
satisfy the modulation equations \re{exactODE epstwof1} up to certain errors, which are expressed in powers of $\eps$ and powers of norms of $v$,$w$.
We introduce a Lyapunov function and compute its time derivative in \cref{Lyapunov functional epstwof}. A lower bound on the Lyapunov function is proved in \cref{chapter Lower bound epsf}. In \cref{Dynamics with approximate
equations for the parameters epstwof}, we prove a version of 
\cref{maintheorem epstwof}
with
approximate
equations for the parameters $(\xi,\u)$.
In \cref{se Pt2 ODE analysis}, we rescale in time the parameters $(\xi,\u)$ and determine 	thereby up to what orders in $\eps$ they differ from exact solutions of ODEs \re{exactODE epstwof1}. The proof of \cref{maintheorem epstwof} is completed in \cref{Completion of the Proof of Theorem}. 
\paragraph{Acknowledgements}
My sincere gratitude goes to my PhD advisor Markus Kunze for the continuous support.
I would like to thank Justin Holmer for helpful discussions.
\paragraph{Notation}
Occasionally we drop the dependence of functions on certain variables. For a Hilbert space $H$ its inner product is denoted by $\langle\cdot,\cdot\rangle_H $.

\section{Symplectic Orthogonal Decomposition}\la{Symplectic Orthogonal Decomposition epsf}
In this section, we show that if $(\t,\p)\in  L^\infty(\R)\oplus L^2(\R)$ is close enough (in the $L^\infty(\R)\oplus L^2(\R)$ norm) to the region 
\be
{\cal S}_0(U):=\l\{ \bma
\t_0(\xi,\u,\cdot)\\
\p_0(\xi,\u,\cdot)
\ema~:~(\xi,\u)\in \Sigma(4,U) 
\r\}\,,
\ee
of the solitary manifold 
${\cal S}_0$, then there exists a unique $(\xi,\u)\in \Sigma(2,U)$ such that we are able to decompose the solution in the following way. 
The soluion can be written as the sum
$$
\bma
\t(\cdot)\\
\p(\cdot)
\ema
= \bma
\t_0(\xi,\u,\cdot)\\
\p_0(\xi,\u,\cdot)
\ema+
\bma
\v(\cdot)\\
\w(\cdot)
\ema\,,
$$
where
$(\t_0(\xi ,\u ,\cdot),\p_0(\xi ,\u ,\cdot))$
is a point on the solitary manifold and $(v( \cdot),w( \cdot))$ is a transversal component, which
is symplectic orthogonal to the tangent vectors 
$
\bma
\dxi\t_0(\xi,\u,\cdot)\\
\dxi\p_0(\xi,\u,\cdot)
\ema
$
and
$
\bma
\du\t_0(\xi,\u,\cdot)\\
\du\p_0(\xi,\u,\cdot)\\
\ema
$
at the corresponding point 
of the solitary manifold ${\cal S}_0$, i.e., the orthogonality condition 
$$
{\cal N} (\t,\p,\xi,\u)=0\,
$$
is satisfied.
In the next lemma, we prove that the symplectic decomposition is  possible in a small uniform distance to the solitary manifold ${\cal S}_0$. 
\ble\la{le uniform decomposition cl}
Let $0<U<1$. 
Let
\be
{\cal O}={\cal O}_{U,p}=\Big\{(\t,\p)\in L^\infty(\R)\times L^2(\R): \inf_{{(\xi,\u)}\in  \Sigma(4,U)} \l| \bma \t(\cdot)\\ \p(\cdot) \ema - \bma\t_0(\xi,\u,\cdot)\\ \p_0(\xi,\u,\cdot)\ema\r|_{L^\infty(\R)\oplus L^2(\R)} <p\Big\}\,.
\ee
There exists $r>0$ such that if $p \le r$  then for any $(\t, \p) \in {\cal O}_{U,p}$ there exists a unique 
$(\xi,\u)\in \Sigma(2,U)$ such that 
$$
{\cal N}(\t,\p,\xi,\u)=0\,
$$
and the map 
$$(\t,\p) \mapsto (\xi(\t,\p) ,\u(\t,\p) )$$ 
is in $C^1({\cal O}_{U,p}, \Sigma(2,U))$.
\ele
\bpr
Notice that
$ U(4) \le U(3)\le U(2)$ and
$\Sigma(4,U)
\subset \Sigma(3,U)\subset \Sigma(2,U)
$.
We consider $(\xi_0, u_0)\in\Sigma(3,U)$. 
Since
\be\la{DxiuNeps} 
{}&D_{\xi,\u} {\cal N}(\t_0(\xi_0,\u_0,x),\p_0(\xi_0,\u_0,x),\xi_0,\u_0)
=
\bma
0
& \g^3(\u_0)m \\
-\g^3(\u_0)m 
& 0
\ema\,, 
\ee
we obtain 
\be\la{detnotzero}
\det D_{\xi,\u} {\cal N}(\t_0(\xi_0,\u_0,x),\p_0(\xi_0,\u_0,x),\xi_0,\u_0)\not=0.
\ee
We prove that there exist $r>0,{\bar\delta}>0$
such that 
$\forall (\xi_0,\u_0)\in \Sigma(3,U)$ there exist balls 
\be
{}& B_r(\t_0(\xi_0,\u_0,\cdot),\p_0(\xi_0,\u_0,\cdot)) \subset L^\infty(\R)\oplus L^2(\R)\,,
~~~B_{\bar\delta}(\xi_0,\u_0)\subset \Sigma(2,U)\,
\ee 
and a map
$$
T_{\xi_0,\u_0}:  B_r(\t_0(\xi_0,\u_0,\cdot),\p_0(\xi_0,\u_0,\cdot)) \to B_{\bar\delta}(\xi_0,\u_0)
$$
such that 
$
T_{\xi_0,\u_0}(\t_0(\xi_0,\u_0,x),\p_0(\xi_0,\u_0,x)) = (\xi_0,\u_0)
$ and
$ 
{\cal N}(\t,\p,T_{\xi_0,\u_0}(\t,\p))=0
$ 
on\\
$B_r(\t_0(\xi_0,\u_0,\cdot),\p_0(\xi_0,\u_0,\cdot))$.  
Therefore we refer to \cite[Theorem 15.1]{Deimling} and check their proof of the implicit function theorem whereas we show that $r$ and ${\bar\delta}$ do not depend on $(\xi_0,\u_0)$. 
We introduce  
$$
\bar{\cal  N}_{\xi_0,\u_0}(\t,\p,\xi,\u)={\cal  N}(\t(\cdot)+\t_0(\xi_0,\u_0,\cdot),\p(\cdot)+\p_0(\xi_0,\u_0,\cdot),\xi+\xi_0,\u+\u_0).$$
It holds 
$\bar{\cal N}_{\xi_0,\u_0}(0,0,0,0)=(0,0).
$
We set
$
K_{\xi_0,\u_0} :=
D_{(\xi,\u)} \bar{\cal N}_{\xi_0,\u_0}(0,0,0,0)
$ and 
\be
S_{\xi_0,\u_0}(\t,\p,\xi,\u)={}& K_{\xi_0,\u_0}^{-1}\bar{\cal N}_{\xi_0,\u_0}(\t,\p,\xi,\u)- I (\xi,\u),
\ee
which is well defined due to \re{detnotzero}.
%
There exists $B>0$ such that
\be
\begin{split}
{}&\forall  (\xi,u)\in\R\times [-U- V(2),U+ V(2)],~\beta_1+\beta_2\le 2,~ p=1,2:\\
{}&\l|\dxi^{\beta_1}\du^{\beta_2}\t_0(\xi,\u,\cdot)\r|_{L^p(\R)}\le B,~~~~~~
\l|\dxi^{\beta_1}\du^{\beta_2}\p_0(\xi,\u,\cdot)\r|_{L^p(\R)}\le B.
\end{split}\nn
\ee
Notice that 
\be
\forall ~(\xi,u)\in \Sigma(2,U):~~~
 \fr 1 {|\g(u)^3 m|} \le \fr 1 c  \,.
\ee
In this proof we denote by $\Vert \cdot\Vert$ the maximum row sum norm of a $2\times 2$ matrix induced by the maximum norm $|\cdot|_{\infty}$ in $\R^2$. We claim that $\exists k\in (0,1),{\bar\delta} >0, ~~\forall    (\xi_0,\u_0)\in \Sigma(3)  
$
$\forall \l((\t,\p),(\xi,\u)\r)\in B_{\bar\delta}(0)\times B_{\bar\delta}(0):~\Vert D_{(\xi,\u)}S_{\xi_0,\u_0} (\t,\p,\xi,\u) \Vert \le k <1$.
Due to \re{DxiuNeps} it holds that

\be
{}&D_{(\xi,\u)}S_{\xi_0,\u_0}(\t,\p,\xi,\u)\\
={}& \fr 1{\gz^3 m} 
\bma
-\dxi\bar{\cal  N}_{\xi_0,\u_0}^2(\t,\p,\xi,\u)  & -\du\bar{\cal  N}_{\xi_0,\u_0}^2(\t,\p,\xi,\u)\\
 \dxi\bar{\cal  N}_{\xi_0,\u_0}^1(\t,\p,\xi,\u) & \du\bar{\cal  N}_{\xi_0,\u_0}^1(\t,\p,\xi,\u)
\ema
- \bma
1 & 0\\
0 & 1
\ema\,.
\ee
The claim follows by estimating each entry of $D_{(\xi,\u)}S_{\xi_0,\u_0} (\t,\p,\xi,\u)$, for instance:
\be
{}&|-\fr 1 {\gz^3 m}\dxi\bar{\cal  N}_{\xi_0,\u_0}^2(\t,\p,\xi,\u)-1|\\
\le{}&
\fr 1 {|\gz^3 m|}\Big(
|\dxi\du\p_0(\bar\xi,\bar\u,x)|_{L^1_x(\R)}|\t(x)+\t_0(\xi_0,\u_0,x)-\t_0(\bar\xi,\bar\u,x)|_{L^\infty_x(\R)}\\
{}&+|\dxi\du\t_0(\bar\xi,\bar\u,x)|_{L^2_x(\R)}|\p(x)|_{L^2_x(\R)}\\
{}&+|\dxi\du\t_0(\bar\xi,\bar\u,x)|_{L^1_x(\R)}|\p_0(\xi_0,\u_0,x)-\p_0(\bar\xi,\bar\u,x)|_{L^\infty_x(\R)}\Big)\\
{}&+|-\fr 1 { \gz^3 m}\int -\du\p_0(\bar\xi,\bar\u,x) \dxi\t_0(\bar\xi,\bar\u,x)+\du\t_0(\bar\xi,\bar\u,x) \dxi\p_0(\bar\xi,\bar\u,x) \,dx-1|.
\ee
Similarly as above one shows that $\exists r\le {\bar\delta} ~~ \forall   (\xi_0,\u_0)\in \Sigma(3) ~  
$
$\forall (\t,\p)\in B_r(0):~|S_{\xi_0,\u_0} (\t,\p,0,0) |_{\infty}< {\bar\delta}(1-k)$,
which completes the proof.

\epr

\section{Existence of Dynamics and the Orthogonal Component} \la{local solution epstwof}
In the following we argue similar to \cite[Proof of theorem 2.1]{Stuart1}. In order to be able to make use of existence theory we 
consider the problem
\be 
\bma
\bar\v(0,x)\\
\bar\w(0,x)
\ema{}& =
\bma
\t(0,x)-\t_0(\xi_s,\u_s,x)\\
\p(0,x)-\p_0(\xi_s,\u_s,x)
\ema\la{vweqn ID epstwof}\,,\\
\dt\bma
\bar\v(t,x)\\
\bar\w(t,x)
\ema
{}&=\bma
\bar\w(t,x)-\p_0(\xi_s,\u_s,x)\\
[\bar\v(t,x)+\t_0(\xi_s,\u_s,x)]_{xx}-\sin (\bar\v(t,x)+\t_0(\xi_s,\u_s,x))+ \eps^2 f(\eps x) \\
\ema\,.
\la{vweqn epstwof}
\ee
By \cite[Theorem VIII 2.1, Theorem VIII 3.2
]{Martin} there exists a local solution (see also \cite[Proof of theorem 2.1]{Stuart1}, \cite[p.434
]{Stuart2}), where
$$
(\bar \v,\bar\w)\in C^1([0,T_{loc}], H^1(\R) \oplus L^2(\R))\,.
$$ 
The function $(\t,\p)$ given by $\t(t,x)=\bar\v(t,x)+\t_0(\xi_s,\u_s,x)$ and $\p(t,x)=\bar\w(t,x)+\p_0(\xi_s,\u_s,x) $ solves obviously locally the Cauchy problem 
\re{SGE1 initialdataprob introduction}
and $(\t,\p) \in C^1([0,T_{loc}], L^\infty(\R) \oplus L^2(\R))$ due to Morrey's embedding theorem. 
We are going to obtain a bound in 
\cref{Dynamics with approximate
equations for the parameters epstwof}
which will imply that the local solution is indeed continuable.
So from now we assume that $(\bar\v,\bar\w) \in C^1([0,\overline T], H^1(\R) \oplus L^2(\R))$ is a solution of \re{vweqn ID epstwof}-\re{vweqn epstwof}	and $(\t,\p)$ is a solution of \re{SGE1 initialdataprob introduction} such that $(\t,\p) \in C^1([0,\overline T], L^\infty(\R) \oplus L^2(\R))$, where $\overline T>0$.

Given $(\t,\p)$ we choose the parameters $(\xi(t),\u(t))$ according to \cref{le uniform decomposition cl} and define $(v,w)$ as follows:
\be
{}&v(t,x)=\t(t,x)-\tztt \la{decomposition1 epstwof}\,,\\
{}&w(t,x)=\p(t,x)-\pztt \la{decomposition2 epstwof}\,.
\ee
$(v(t,x),w(t,x))$ is well defined for $t \ge 0$ so small that 
$
\nlinf{v(t)}+\nltwo{w(t)}\le r\,
$
and
$
(\xi(t),\u(t))\in \Sigma(4,U) \,,
$
where $r$ and $U$ are from \cref{le uniform decomposition cl}. 
We formalize this in the following definition.

\bde\la{de: exittime decomposition epstwof}
Let $t^*$ be the "exit time":
\be
t^*:= \sup\Big\{{}& T>0:\tnvinf{v}{0}{t}+\tnw{w}{0}{t}\le {r },\\ 
{}& (\xi(t),\u(t))\in \Sigma(4,U),~0\le t \le T\Big\}\,,
\ee
where $r$ and $U$ are from \cref{le uniform decomposition cl}. 
\ede
\noindent
Notice that $(\xi_s,\u_s)=(\xi(0),\u(0))\in \Sigma(4,U) $. We will choose $\eps$ such that, among others,  
$$\nlinf{v(0)}+\nltwo{w(0)}\le \fr r 2,$$ where $(\v(0),\w(0))$ is given by \re{SGE1 initialdataprob introduction}.
Thus $(v(t,x),w(t,x))$ is well defined for $0\le t \le t^*$.
In the following lemma we obtain more information on $(v,w)$.
\ble
Let $T=\min\{ t^*, \overline T\}$ and let  
$(v,w)$ be defined by \re{decomposition1 epstwof}-\re{decomposition2 epstwof}. Then
$(\v, \w) \in C^1([0, T], H^1(\R) \oplus L^2(\R))$.
\ele
\bpr
This follows by using \re{decomposition1 epstwof}-\re{decomposition2 epstwof} and the fact that $(\bar\v,\bar\w) \in C^1([0,\overline T], H^1(\R) \oplus L^2(\R))$,
since the difference $( \t_K(\g(u_0)(\cdot-\xi_0))-\t_K(\g(\bar \u)(\cdot-\bar\xi)) )$ is in $ L^2(\R)$ for all $(\xi_0,u_0),(\bar\xi,\bar u)\in \R\times (-1,1)$.
\epr

\noindent
We compute the time derivatives of $v$ and $w$, which will be needed in the following sections.
\ble\la{dvdw epstwo}
The equations for $(v,w)$, defined by \re{decomposition1 epstwof}-\re{decomposition2 epstwof}, are
\be
\d \v(x) = {}&\w(x)- \d\xi\dxi\tzt -\d\u\du\tzt+\u\dxi\tzt \,,\\ 
\d \w(x)  ={}&\dx^2\v(x)-\cos\tzt \v(x)+\eps^2 f(\eps x)+\fr{\sin\tzt\v^2(x)}{2}+\tilde R(\v)(x)\\
{}& +\u\dxi\pzt- \d\xi\dxi\pzt -\d\u\du\pzt\,, 
\ee
for times $t \in [0,t^*]$, where $\tilde R(\v)={\cal O}(\nhone{v}^3)$.

\ele
\bpr
By taking the time derivatives of $(v,w)$ and using \re{decomposition1 epstwof}-\re{decomposition2 epstwof},
\re{SGE1 initialdataprob introduction} we obtain
\be
\d \v(x) ={}& \w(x)+\pzt 
- \d\xi\dxi\tzt -\d\u\du\tzt \\
= {}&\w(x)- \d\xi\dxi\tzt -\d\u\du\tzt+\u\dxi\tzt \,   
\ee
and
\be
\d \w(x) ={}& \dx^2\t(x)-\sin\t(x)+\eps^2 f(\eps x)
- \d\xi\dxi\pzt -\d\u\du\pzt \\
={}&
\dx^2\tzt+\dx^2\v(x)-\sin\tzt
-\cos\tzt\v(x)+\fr{\sin\tzt\v^2(x)}{2}\\
{}&+\tilde R(\v)(x)+ \eps^2 f(\eps x)
- \d\xi\dxi\pzt -\d\u\du\pzt\\
{}&+\u\dx\pzt-\u\dx\pzt\\
={}&\dx^2\v(x)-\cos\tzt \v(x)+\eps^2 f(\eps x)+\fr{\sin\tzt\v^2(x)}{2}+\tilde R(\v)(x)\\
{}& +\u\dxi\pzt- \d\xi\dxi\pzt -\d\u\du\pzt \,, 
\ee
where we have expanded the term $\sin(\tzt+\v(x))$.
\epr
\section{Modulation Equations}\la{Modulation Equations for ODE analysis epstwof}

The restriction of the Hamiltonian $H^\eps$  given by \re{intro Hamiltonian}, with $F(\eps,x)=\eps^2 f(\eps x)$, to the solitary manifold ${\cal S}_0$, can be expressed in the form 
\be
 H^\eps(\t_0(\xi,\u,x),
\p_0(\xi,\u,x)) 
=   m  \g(u) - \int  \eps^2 f(\eps ( y+\xi)) \t_K(\g(\u)y) \,dx 
\ee
for an appropriate $f$.
The derivatives with respect to $u$ and $\xi$ are given by
\be
\du H^\eps(\t_0(\xi,\u,x),
\p_0(\xi,\u,x)) 
={}&  m  \u [\g(u)]^3
 - \u [\g(u)]^3 \int  \eps^2  f(\eps ( y+\xi)) y\t_K'(\g(\u)y) \,dy,  
\\	
	\dxi H^\eps(\t_0(\xi,\u,x),
\p_0(\xi,\u,x))  
={}&   \g(\u)  \int  \eps^2 f(\eps ( y+\xi)) \t_K'(\g(\u)y) \,dx.
\ee
Thus we obtain for the restricted Hamiltonian
$
 (\xi,\u)\mapsto H^\eps(\t_0(\xi,\u,x),\p_0(\xi,\u,x)) 
$
the following Hamiltonian equations of motion (with respect to the corresponding restricted symplectic form) 
\be
 \d \xi  
={}&   \u 
-   \eps^3\fr{ f'(\eps \xi)\u}{{[\g(u)]^3 m}} \int Z^2 \t_K ' (Z) \,dZ +{\cal O}(\eps ^5)\la{restricted eq 1},
\\
 -  \d u 
={}& \eps^2\fr{ f(\eps \xi)}{[\g(u)]^3 m}\int \t_K ' (Z) \,dZ
+{\cal O}(\eps ^4)
\la{restricted eq 2}
 \,,
\ee
where we expanded $f(\eps ( y+\xi))$ around $y=0$ and used that $Z\t_K'(Z)$ is an odd function.
In the following we derive estimates which show how far the parameters $(\xi(t),\u(t))$ satisfy the equation above.
Let us start with a definition.
\bde
Let $\eps>0$, $(\xi,\u)\in \R\times (-1,1)$. We set 
$$ W(\eps,\xi,\u) := \fr{\eps^2 f(\eps \xi)\int \t_K ' (Z) \,dZ }{[\g(u)]^3 m} \,.$$ 
\ede

\ble\la{le Modulation Equations epstwof}
There exists an $\eps_0>0$ such that the following statement holds.
Let $\eps\in(0,\eps_0]$ and $(v,w)$ be given by \re{decomposition1 epstwof}-\re{decomposition2 epstwof}, with $(\xi,\u)$ obtained from \cref{le uniform decomposition cl}. Let 
$$ \tnv{v}{0}{t^*},\tnw{w}{0}{t^*}\le \eps_0\,, $$
where $t^*$ is from \cref{de: exittime decomposition epstwof}. Then
\be
|\d\xi -u |  \le{}&  C[\nhone{v}+\nltwo{w}] \eps^2 +C \eps^3+C\nhone{v}^2 ,\\
	|\d\u + W(\eps,\xi,\u) | \le{}&  C[\nhone{v}+\nltwo{w}] \eps^2 +C \eps^4+C\nhone{v}^2 ,
\ee
for $0\le t \le t^*$, where $C$ depends on $f$.
\ele
\bpr
The technique we use is similar to that in the proof of  \cite[Lemma 6.2]{ImaykinKomechVainberg}.
We start with some definitions and set
\be
\Omega  ( \u):=
\bma
\OM{t_{1 } (\xi,\u,\cdot)}{t_{1 } (\xi,\u,\cdot)} & \OM{t_{1 } (\xi,\u,\cdot)}{t_{2 } (\xi,\u,\cdot)}\\
\OM{t_{2 } (\xi,\u,\cdot)}{t_{1 } (\xi,\u,\cdot)} & \OM{t_{2 } (\xi,\u,\cdot)}{t_{2 } (\xi,\u,\cdot)}\\
\ema\nn
%
=
\g(\u)^3 m  \bma
 0 & 1\\
-1				& 0 
\ema.\nn
\ee 
Now we consider for any $(\bar\xi,\bar u)\in\R\times [-U- V(2),U+ V(2)]$, $(\bar v,\bar w)\in H^1(\R)\times L^2(\R)$ the matrix:
\be
{}&M (\bar\xi,\bar\u,\bar\v,\bar\w)\nn\\
={}&\bma
\Ltwo{\bma \dxi^2\p_0(\bar\xi,\bar\u,\cdot) \\ -\dxi^2\t_0(\bar\xi,\bar\u,\cdot) \ema}{\bma \bar\v(\cdot)\\ \bar\w(\cdot) \ema}
 & \Ltwo{\bma \du\dxi\p_0(\bar\xi,\bar\u,\cdot)\\ -\du\dxi\t_0(\bar\xi,\bar\u,\cdot) \ema}{\bma \bar\v(\cdot) \\ \bar w(\cdot) \ema} \\
\Ltwo{\bma \dxi\du\p_0(\bar\xi,\bar\u,\cdot)\\ -\dxi\du\t_0(\bar\xi,\bar\u,\cdot) \ema}{\bma \bar\v(\cdot) \\ \bar w(\cdot) \ema}  &  \Ltwo{\bma \du^2 \p_0(\bar\xi,\bar\u,\cdot)\\ -\du^2\t_0(\bar\xi,\bar\u,\cdot) \ema}{\bma \bar\v(\cdot) \\ \bar w(\cdot) \ema}
\ema.\nn
\ee
It holds for all $(\bar\xi, \bar u)\in\R\times [-U- V(2),U+V(2)]$, $(\bar v,\bar w)\in H^1(\R)\times L^2(\R)$:
\be\la{M small v w}
 \Vert \l[\Omega  ( \bar\u)\r]^{-1}  M (\bar\xi,\bar\u,\bar\v,\bar\w)\Vert \le C (\nhone{\bar\v}+ \nltwo{\bar\w}),
\ee
where we denote by $\Vert \cdot \Vert$ a matrix norm. Let $I=I_2$ be the identity matrix of dimension 2. Due to \re{M small v w} we are able to choose $\eps_0>0$ such that if $ \nhone{\bar v},\nltwo{\bar w}\le \eps_0$ then the matrix
$$
I+\l[\Omega  ( \bar\u)\r]^{-1}  M (\bar\xi,\bar\u,\bar\v,\bar\w)
$$
is invertible by von Neumann's theorem. 
Using
\re{decomposition1 epstwof}-\re{decomposition2 epstwof} we express the orthogonality condition ${\cal N}(\t,\p,\xi,\u)=0$ from \cref{le uniform decomposition cl} in terms of $(\v,\w,\xi,\u)$ 
and take its derivative with respect to $t$.
For simplicity of notation, we drop $(\t,\p,\xi,\u)$ and obtain using \cref{dvdw epstwo}
in matrix form:
\be
0{}&=\ddt\bma
{{\cal N}_1}\\
{{\cal N}_2}\\
\ema
=\Omega
\bma
\d\xi-u\\
\d\u+W(\eps,\xi,\u)\\
\ema
+ M \bma
\d\xi-u\\
\d\u+W(\eps,\xi,\u)\\
\ema
+ \bma
P_1\\
P_2
\ema,\nn
\ee
where 
$M=M (\xi,\u,\v,\w)$, $\Omega=\Omega  (\u)$, $P_1=P_{1 } (\xi,\u,\v,\w)$, $P_2 =P_{2 } (\xi,\u,\v,\w)$,
\be
P_1(\xi,\u,\v,\w)=
{}&\Ltwo{\bma
-\u\dx\v(\cdot)-\w(\cdot) \\ 
-\dx^2\v(\cdot)+\cos(\t_0(\xi,\u,\cdot))\v(\cdot)-\u\dx\w(\cdot) \\ 
\ema}{ 
\bma
-\dxi\p_0(\xi,\u,\cdot)\\
\dxi\t_0(\xi,\u,\cdot)\\
\ema}\\
{}&{-\int
\dxi\tzt
\eps^2 f(\eps x)  
\,dx-W(\eps,\xi,\u)\g(u)^3m}\\
{}&{-\int\du\dxi\pzt\v(x)-\du\dxi\tzt\w(x) \,dx\cdot W(\eps,\xi,\u)}\\
{}&{  -\int\dxi\tzt
\fr{\sin\tzt\v^2(x)}{2}
+\tilde R(\v)(x)
\,dx} 
%
\ee
and
\be
P_2(\xi,\u,\v,\w) =
{}&\Ltwo{\bma
-\u\dx\v(\cdot)-\w(\cdot) \\ 
-\dx^2\v(\cdot)+\cos(\t_0(\xi,\u,\cdot))\v(\cdot)-\u\dx\w(\cdot) \\ 
\ema}{ 
\bma
-\du\p_0(\xi,\u,\cdot)\\
\du\t_0(\xi,\u,\cdot)\\
\ema}\\
{}&-\int
\du\tzt 
\eps^2 f(\eps x)
\,dx\\
{}&-\int \du^2\dxi\pzt\v(x)-\du^2\tzt\w(x)\,dx \cdot W(\eps,\xi,\u) \\
{}&-\int
\du\tzt 
\fr{\sin\tzt \v^2(x)}{2}+\tilde R(v)(x)
\,dx\,. 
\ee
If $ \nhone{v},\nltwo{ w}\le \eps_0$ then we obtain as mentioned above by von Neumann's theorem that 
\be
\bma
\d\xi-u\\
\d\u+W(\eps,\xi,\u)\\
\ema=
-\Big( I+ \Omega^{-1}M \Big)^{-1} [\Omega^{-1}\bma P_1 \\ P_2 \ema]\,.
\ee
Now we take a closer look at the terms that occur in $P_1$ and $P_2$.  Integration by parts
and symplectic orthogonality yield that   
\be
{}&\Ltwo{\bma
-\u\dx\v(\cdot)-\w(\cdot) \\ 
-\dx^2\v(\cdot)+\cos(\t_0(\xi,\u,\cdot))\v(\cdot)-\u\dx\w(\cdot) \\ 
\ema}{ 
\bma
-\dxi\p_0(\xi,\u,\cdot)\\
\dxi\t_0(\xi,\u,\cdot)\\
\ema}=0\,,\\
{}&\Ltwo{\bma
-\u\dx\v(\cdot)-\w(\cdot) \\ 
-\dx^2\v(\cdot)+\cos(\t_0(\xi,\u,\cdot))\v(\cdot)-\u\dx\w(\cdot) \\ 
\ema}{ 
\bma
-\du\p_0(\xi,\u,\cdot)\\
\du\t_0(\xi,\u,\cdot)\\
\ema}=0\,,
\ee
which can also be deduced from \cite[Lemma A.5]{Mashkin}.
It holds that 
\be
{}&{-\int
\dxi\tzt
\eps^2 f(\eps x)  
\,dx-W(\eps,\xi,\u)\g(u)^3m}\\
={}&\int\t_K'(Z)\eps^2 f(\eps (\fr Z {\g(u)}+\xi))\,dZ-\eps^2 f(\eps \xi)\int \t_K ' (Z) \,dZ\\
={}&\int\t_K'(Z)\eps^2 \Big[f(\eps \xi)+ \fr \eps{\g(u)} f'(\eps\xi)Z+\ldots\Big]\,dZ-\eps^2 f(\eps \xi)\int \t_K ' (Z) \,dZ\, 
\ee
and
\be
{}&
-\int
\du\tzt 
\eps^2 f(\eps x)
\,dx
\\
=
{}&-\u\g(u)\int Z\t_K'(Z)
\eps^2 f(\eps (\fr Z{\g(u)}+\xi))
\,dZ\\
=
{}&-\u\g(u)\int Z\t_K'(Z)
\eps^2 \Big[f(\eps \xi)+ \fr \eps{\g(u)} f'(\eps\xi)Z+\ldots\Big]
\,dZ\,.
\ee
Since $Z\t_K'(Z)$ is an odd function we obtain 
\be
|P_1|\le 
C[\nhone{v}+\nltwo{w}] \eps^2 +C \eps^4+C\nhone{v}^2\,,
\\
|P_2| \le 
C[\nhone{v}+\nltwo{w}] \eps^2 +C \eps^3+C\nhone{v}^2\,.
\ee
\epr
\section{Lyapunov Function}\la{Lyapunov functional epstwof}
In this section we introduce the Lyapunov function and compute its time derivative. 
\bde
Let $(v,w)$ be given by \re{decomposition1 epstwof}-\re{decomposition2 epstwof}, with $(\xi,\u)$ obtained from \cref{le uniform decomposition cl}.
We define the Lyapunov function $ L $ by 
\be 
\la{lyapunovfunction epstwof}
{}& L =\int \fr{\w^2(x)} 2 +\fr{\dx\v^2(x)} 2 +\fr{\cos(\t_K(\g(u)(x-\xi))) \v^2(x)}2+\u\w(x)\dx\v(x)\,dx\,.
\ee
\ede

\ble\la{lelyapunovfunction epstwo}
It holds for times $t \in [0,t^*]$ that 
\be
\ddt { L}  
=
{}&\int\w(x)\l[\fr{\sin(\tzt)\v^2(x)}{2}+\ti R(\v)(x) \r]\\
{}&+\u\dx\v(x)\l[\fr{\sin(\tzt)\v^2(x)}{2}+\ti R(\v)(x) \r]\,dx\\
{}&-\d\u\int \fr{\sin(\tzt)}2\du\tzt\v^2(x)\,dx\\
{}&+ (\d\xi-\u) \int  {\cos(\tzt)} \v(x)\dx\v(x)\,dx+\d\u\int \w(x)\dx\v(x)\,dx\\
{}&+ \eps^2 \int \d\v(x)f(\eps x)\,dx 
-\u\d u \g(u)^3\eps^2\int (x-\xi)\t_K'(\g(u)(x-\xi))f(\eps x)\,dx\\ 
{}&+(\u-\d\xi)\g(u)\eps^2\int \t_K'(\g(u)(x-\xi)) f(\eps x)\,dx
-\u\eps^3\int \v f'(\eps x)\,dx\,.
\ee
\ele
\bpr
We use a similar technique as in the proof of 
\cite[Lemma 2.1]{KoSpKu}.
We can assume that the initial data of our problem has compact support. This allows us to do the following computations (integration by parts etc.). The claim for non-compactly supported initial data follows by density arguments. 
Since $\int \dx\v(x)\dx^2\v(x)+\w(x)\dx\w(x)\,dx=0$ and
\be
{}&
\int\fr{\dt[\cos\tzt]}2\v^2(x)\,dx
\\
={}&\int \d\xi {\cos(\tzt)}\v(x)\dx\v(x)-\d\u\fr{\sin(\tzt)}2\du\tzt\v^2(x)\,dx\,,
\ee
we obtain by
taking the time derivative of the Lyapunov function \re{lyapunovfunction epstwof} and by using \cref{dvdw epstwo}:
\be
\d L  
={}& (\u-\d\xi)\Ltwo{\bma
-\u\dx\v(\cdot)-\w(\cdot) \\ 
-\dx^2\v(\cdot)+\cos(\t_0(\xi,\u,\cdot))\v(\cdot)-\u\dx\w(\cdot) \\ 
\ema}{ 
\bma
-\dxi\p_0(\xi,\u,\cdot)\\
\dxi\t_0(\xi,\u,\cdot)\\
\ema}\\
{}&-\d\u\Ltwo{\bma
-\u\dx\v(\cdot)-\w(\cdot) \\ 
-\dx^2\v(\cdot)+\cos(\t_0(\xi,\u,\cdot))\v(\cdot)-\u\dx\w(\cdot) \\ 
\ema}{ 
\bma
-\du\p_0(\xi,\u,\cdot)\\
\du\t_0(\xi,\u,\cdot)\\
\ema}\\
{}&+\w(x)\l[\fr{\sin\tzt\v^2(x)}{2}+\ti R(\v)(x) \r]
+\u\dx\v(x)\l[\fr{\sin\tzt\v^2(x)}{2}+\ti R(\v)(x) \r]\\
{}&-\d\u\int \fr{\sin(\tzt)}2\du\tzt\v^2(x)\,dx
+ (\d\xi-\u) \int  {\cos(\tzt)} \v(x)\dx\v(x)\,dx\\
 {}&+\d\u\int \w(x)\dx\v(x)\,dx
+\int \w(x)\eps^2 f(\eps x)\,dx
+\int \u\dx\v(x)\eps^2 f(\eps x)\,dx\,.
\ee
The first two terms vanish by the same argument as in the proof of \cref{le Modulation Equations epstwof}.
Using the identity
\be
\g(u)\u\t_K'(\g(u)(x-\xi))=-\dt[\t_K(\g(u)(x-\xi))]+\l[-\u\d u \g(u)^3(x-\xi)+(\u-\d\xi)\g(u)\r]\t_K'(\g(u)(x-\xi))\,,
\ee
it follows from \re{SGE1 initialdataprob introduction} and \re{decomposition1 epstwof}-\re{decomposition2 epstwof}
that 
\be\la{wfterm}
{}&  \int \w \eps^2f(\eps x)\,dx\\
= {}& \eps^2 \int \d\v(x)f(\eps x)\,dx 
-\u\d u \g(u)^3\eps^2\int (x-\xi)\t_K'(\g(u)(x-\xi))f(\eps x)\,dx\\ 
{}&+(\u-\d\xi)\g(u)\eps^2\int \t_K'(\g(u)(x-\xi)) f(\eps x)\,dx\,. 
\ee
The claim follows, since
\be
\int \u\dx\v(x)\eps^2 f(\eps x)\,dx {}&= -\u\eps^3\int \v(x) f'(\eps x)\,dx \,.
\ee
\epr

\section{Lower Bound}\la{chapter Lower bound epsf}
We introduce a functional ${\cal  E}$ and prove a lower bound on ${\cal  E}$ by using symplectic orthogonality combined with functional analytic arguments. 
This will imply a lower bound on the Lyapunov function $L$,
which will play a key role in the proof of the main result.

\bde
For $(v,w)\in H^1(\R)\times L^2(\R)$,  $(\xi,u)\in\R\times (-1,1)$ we set
$$
{\cal  E}(\v,\w,\xi,\u):=
\fr 1 2 \int(\w(x)+\u\dx\v(x))^2+\v_Z^2(x)+\cos(\t_K(Z))\v^2(x) \,dx
\,,
$$
where $Z=\g(u)(x-\xi)$ and $\v_Z(x)=\dZ\v(\fr Z {\g(u)} +\xi)=\fr 1 {\g(u)} \dx\v(x)$.
\ede
\noindent
A straightforward computation yields the following lemma.   
\ble \la{le Efunctional equal Lfunctional}
For $(v,w)\in H^1(\R)\times L^2(\R)$,  $(\xi,u)\in\R\times (-1,1)$ it holds that
\be
{\cal  E}(\v,\w,\xi,\u)=
\int \fr{\w^2(x)} 2 +\fr{(\dx\v(x))^2} 2 +\fr{\cos(\t_K(\g(u)(x-\xi))) \v^2(x)}2+\u\w(x)\dx\v(x)\,dx.\nn
\ee
\ele
\noindent
Recalling the relations \re{decomposition1 epstwof}-\re{decomposition2 epstwof}  we introduce a notation in order to be able to express the orthogonality conditions in terms of the variables $(\v,\w,\xi,\u)$ instead of the variables 
$(\t,\p,\xi,\u)$.
\bde
For $(v,w)\in H^1(\R)\times L^2(\R)$,  $(\xi,u)\in\R\times (-1,1)$ we set
\be 
{}&{\cal \check N}_1 (\v,\w,\xi,\u)=\int \dxi\p_0 (\xi,u,x)\v(x) - \dxi\t_0(\xi,u,x)\w(x)\,dx,\nn
\\
{}&{\cal \check N}_2 (\v,\w,\xi,\u)=\int \du\p_0(\xi,u,x) \v(x) - \du\t_0(\xi,u,x)\w(x)\,dx.\nn
\ee
\ede
\noindent 
Now we prove a lower bound on the functional ${\cal  E}$. %
\ble[Stuart] \la{leStuart epsf} 
There exists $c>0$ such that if $(\xi,\u)\in \R \times [-U- V(2),U+ V(2)]\subset \R\times (-1,1)$ and $(v,w)\in H^1(\R)\times L^2(\R)$
satisfies 
$${\cal \check N}_2(\v,\w,\xi,\u)=0$$ 
then
$$
{\cal  E}(\v,\w,\xi,\u)
\ge c (\nhone{ v}^2 + \nltwo{ w}^2)\,.
$$
\ele
\bpr We follow closely \cite{Stuart3} and \cite{Stuart1}. This proof is a slight modification of the proof of \cite[Lemma 4.3]{Stuart3}.
Notice that the operator $-\dZ^2+\cos\t_K(Z)$ is nonnegative. It has (see \cite{Stuart2}) an one
dimensional null space spanned by $\t_K'(\cdot)$ and the essential spectrum $[1,\infty)$. 
We argue by contradiction and asume first: 
$\exists \xi \in \R ~~\forall j\in\N~ \exists \u_j\in [-U- V(2),U+ V(2)]~~
 \exists
(\bar v_j,\bar w_j)\in H^1(\R)\times L^2(\R) :$
\be \la{notEnergy epsf} \begin{split} 
 {}&{\cal \check N}_2(\bar\v_j,\bar\w_j,\xi,\u_j)=0\,,~~~~
{\cal  E}(\bar\v_j,\bar\w_j,\xi,\u_j)< \fr 1 j (\nhone{ \bar v_j}^2 + \nltwo{ \bar w_j}^2)\,.
\end{split} \ee 
This statement is also true for the sequences 
$ v_j:=  {\bar v_j} {(\nhone{\bar v_j}^2 + \nltwo{\bar w_j}^2)^{-\fr 1 2}}$ and $w_j:= {\bar w_j} {(\nhone{\bar v_j}^2 + \nltwo{\bar w_j}^2)^{-\fr 1 2}}$. 
Assuming that $\nltwo{ v_j}\OT{\to}{j\to\infty} 0$ we obtain
$\nltwo{ ( v_j)_x}\OT{\to}{j\to\infty} 0$ and 
$\nltwo{  w_j}\OT{\to}{j\to\infty} 0$. This is a contradiction to the fact that $\nhone{ v_j}^2 + \nltwo{  w_j}^2=1~\forall j\in\N$. By passing to a subsequence we may assume (without loss of generality) that there exists ${\bar\delta}>0$ such that 
\be\la{normvj epsf}\nltwo{  v_{j}}^2\ge {\bar\delta} ~\forall j\in\N\,.\ee 
Since $(v_{j}, w_{j})$ is bounded in $H^1(\R)\times L^2(\R)$ we may assume that $ v_j \OT{\rightharpoonup}{H^1(\R)} v$ and $ w_j \OT{\rightharpoonup}{L^2(\R)} w$  by taking subsequences. 
Due to Rellich's theorem we may assume by passing to subsequences again that $ v_j \OT{\rightarrow}{L^2(\Omega)} v$, where $\Omega\subset\R$ is bounded and open. Passing to a further subsequence we may assume almost everywhere convergence
and also $\u_j \OT{\to}{\R}  u$. 
Due to the fact that 
\be\la{rcosinus epsf}
\exists~ r>0~~~\text{s.t.}~~~ |\cos(\t_K(Z))|>\fr 1 2 ~~~\text{for}~~~ |Z|>r
\ee 
and that $-\dZ^2+\cos\t_K(Z)$ is a nonnegative operator we obtain the estimate
\be
{\cal  E}(\v_j,\w_j,\xi,\u_j)
\ge \fr {1} 4 \int_{-\infty}^{\fr {-r} {\g(u_j)}+\xi} \v_j^2(x) \,dx
+\fr {1} 4 \int_{\fr {r} {\g(u_j)}+\xi}^{\infty} \v_j^2(x) \,dx\,,
\ee
where we used integration by parts and substitution.
Hence \re{notEnergy epsf} implies that 
$$
\int_{\{x\in\R:|x|\ge \ti r\}} \v_j^2(x) \,dx \OT{\to}{j\to\infty} 0
$$
for a sufficiently large $\ti r$. As a consequence \re{normvj epsf} and the strong convergence on bounded domains yield 
$$\int_{\{x\in\R:|x| \le \ti r\}} v^2(x)\,dx\ge{\bar\delta},$$ from which it follows that $v \not\equiv 0$. Weak convergence implies using the triangle inequality that
\be
{\cal \check N}_2(\v,\w,\xi,\u)={}&0\la{orth2eqzero epsf}   
\ee
and
\ben
\la{Energyzero1 epsf}
\fr 1 2 \int\l(\w(x)+\u\v'(x)\r)^2\,dx
\le{}&\liminf_{j\to\infty}\fr 1 2 \int\l(\w_j(x)+\u_j\v_j'(x)\r)^2\,dx,\\
\fr 1 2 \int\l(\fr 1 {\g(u)}\v'(x)\r)^2\,dx
\le{}&\liminf_{j\to\infty}\fr 1 2 \int\l(\fr 1 {\g(u_j)}(\v_j)'(x)\r)^2\,dx.
\een
Due to \re{rcosinus epsf} we are able to apply Fatou's lemma for a sufficiently large $\ti r$ and obtain
\ben
\la{Energyzero2}
\begin{split+}
{}&\fr 1 2 \int_{\{x\in\R:|x| > \ti r\}} \cos(\t_K(\g(\u)(x-\xi)))\v^2(x)\,dx\\
\le{}&\liminf_{j\to\infty}\fr 1 2 \int_{\{x\in\R:|x| > \ti r\}}\cos(\t_K(\g(\u_j)(x-\xi_j)))\v_j^2(x)\,dx,
\end{split+}
\een
where we have used that $(v_j)$ converges almost everywhere.
The dominated convergence theorem yields
\be\la{Energyzero3 epsf}\begin{split}
{}&\fr 1 2 \int_{\{x\in\R:|x| \le \ti r\}} \cos(\t_K(\g(\u)(x-\xi)))\v^2(x)\,dx\\
={}&\lim_{j\to\infty}\fr 1 2 \int_{\{x\in\R:|x| \le \ti r\}}\cos(\t_K(\g(\u_j)(x-\xi_j)))\v_j^2(x)\,dx\,.
\end{split}
\ee
\re{notEnergy epsf} together with \re{Energyzero1 epsf}-\re{Energyzero3 epsf} imply that
$
{\cal  E}(\v,\w,\xi,\u)=0
$. It follows that
$(v(x),w(x))=\a(\t_K'(\g(u)(x-\xi)),-\u\g(u)\t_K''(\g(u)(x-\xi)))$ for some $\a\not=0$, since $v \not\equiv 0$. This is a contradiction to \re{orth2eqzero epsf}. 
The constant $c$ does not depend on $\xi$, since
$
\check{\cal N}_2(\v,\w,\xi,\u)=\check{\cal N}_2(\v(\cdot+\xi),\w(\cdot+\xi),0,\u)=0
$
implies that
$
{\cal  E}(\v,\w,\xi,\u)={\cal  E}(\v(\cdot+\xi),\w(\cdot+\xi),0,\u)\ge c(0)(\nhone{ v}^2 + \nltwo{w}^2)
$.
\epr

\bre
Let $(v,w)$ be given by \re{decomposition1 epstwof}-\re{decomposition2 epstwof}, with $(\xi,\u)$ obtained from \cref{le uniform decomposition cl}. It holds that
$
L(t)={\cal  E}(\v(t),\w(t),\xi(t),\u(t))\,
$
for times $t \in [0,t^*]$.
\ere

\section{Description of the Dynamics with Approximate
Equations for the Parameters $(\xi,\u)$}\la{Dynamics with approximate
equations for the parameters epstwof}
We prove first a version of \cref{maintheorem epstwof} with approximate equations for the parameters $(\xi,\u)$. 
\bth\la{th: preparing maintheorem epstwof}
Suppose that the assumptions 
of \cref{maintheorem epstwof}
are satisfied.
Then the Cauchy problem defined by \re{SGE1 initialdataprob introduction}
has a unique solution on the time interval
\be
0\le t \le T  ~\text{where}~ 
T=T(\eps)=
\fr {1} {\eps^{ }}.
\ee
The solution may be written in the form
\be
\bma
\t(t,x)\\
\p(t,x)
\ema
=
\bma
\tztt\\
\pztt
\ema
+
\bma
\v(t,x)\\
\w(t,x)
\ema
\ee
where $\v, \w, \u, \xi$ have regularity
$
(\xi(t), \u(t)) \in C^1([0,T ] ,\R \times (-1, 1)) 
$, 
$(v(t), w(t)) \in C^1([0,T ], H^1(\R) \times L^2(\R))$
such that the orthogonality condition  
$${\cal N}(\t(t,x),\p(t,x),\xi(t),\u(t))=0\,$$
is satisfied. There exist positive constants $c,C$ such that
\be
|\d\xi -u |  
\le   C \eps^{\fr {11} 4} ,~~~~ ~~
	|\d\u  + W(\eps,\xi,\u) | 
	\
\le   C \eps^{\fr {11} 4} \, , 
\ee
and
\be
\tnv{v}{0}{T}^2+\tnw{w}{0}{T}^2\le c \eps^{\fr {11} 4}\,.
\ee
The constants $c,C$ depend on $f$.
\eth
\noindent
\cref{th: preparing maintheorem epstwof} yields approximate equations for the parameters $(\xi,\u)$ whereas \cref{maintheorem epstwof} provides ODEs \re{exactODE epstwof1} which are exactly satisfied.
The bound for the transversal component $(\v,\w)$ in \cref{th: preparing maintheorem epstwof} is better than in \cref{maintheorem epstwof}. Notice further that in \cref{th: preparing maintheorem epstwof} 
the orthogonality conditions are satisfied which do not have to hold in \cref{maintheorem epstwof}.\\

The proof of \cref{th: preparing maintheorem epstwof} needs some preparation. 
Now we suppose that \re{SGE1 initialdataprob introduction} has a solution and we make some assumptions on $(v,w)$ given by \re{decomposition1 epstwof}-\re{decomposition2 epstwof} and on $(\xi,\u)$ obtained from \cref{le uniform decomposition cl}. Then the  following lemma yields us more accurate information about $(v,w)$ and $(\xi,\u)$.

\ble\la{le: prepairing main proof epstwof}
Suppose that the assumptions 
of \cref{maintheorem epstwof} on $f$ and $(\xi_s,u_s)$
are satisfied and let $\eps$ be  sufficiently small.
Assume that \re{SGE1 initialdataprob introduction} has a solution $(\t,\p)$ on $[0,\overline T]$ such that
$$
(\t,\p) \in C^1([0,\overline T], L^\infty(\R) \oplus L^2(\R))\,.
$$
Suppose that
$
0\le T \le t^*\le \overline T\,
$
with $t^*$ from \cref{de: exittime decomposition epstwof}.
Let $(v,w)$ be given by \re{decomposition1 epstwof}-\re{decomposition2 epstwof}, with $(\xi,\u)$ obtained from \cref{le uniform decomposition cl}
such that
$$\tnv{v}{0}{T}^2+\tnw{w}{0}{T}^2\le  \eps^{5/2}\,.$$
Then, provided
$
0\le  T\le  {\eps^{ -1 }},
$
it holds that
\begin{itemize}
\item[(a)] $\forall  t \in [0,T]:~(\xi(t),u(t))\in \Sigma(5,U)$;
\item[(b)] $\tnv{v}{0}{T}^2+\tnw{w}{0}{T}^2\le \bar C (L(0)+\eps^{11/4})\,,$
where $\bar C$ depends on $f$ (and on $c$ from \cref{leStuart epsf}).
\end{itemize}
\ele
\bpr
%
\cref{le Modulation Equations epstwof} yields for times $t \in [0,T]$:
\be
|\d\xi -u |  
\le{}&   C[\nhone{v}+\nltwo{w}] \eps^2 +C \eps^3+C\nhone{v}^2 
\le   C \eps^{5/2 } ,\\
	|\d\u + W(\eps,\xi,\u) | \le {}&  C[\nhone{v}+\nltwo{w}] \eps^2 +C \eps^4+C\nhone{v}^2\,  
\le   C \eps^{5/2} \,. 
\ee
Thus we obtain for times $  t \in [0,T]$:
$
 |u(t)-u(0)|
\le  \int _0^t |\d u(s)| \,ds  
\le  C \eps^{2} t  
$,
which implies
$ |u(t)| \le  C \eps^{2} t + |u(0)| $.
It follows (a) due to $|u_s|<U$ and the smallness assumption on $\eps$.
Using \cref{lelyapunovfunction epstwo}, \cref{le Efunctional equal Lfunctional} and \cref{leStuart epsf} we obtain for times
$
0\le t\le T\le {\eps^{-1 }}, 
$
the following estimate,
\be
{}&{\CR c}(\nhone{v(t)}^2+\nltwo{w(t)}^2)\\
\le {}&  L(t) 
=  L(0)
+\int_0^t \d{L}(t)\,dt\\
\le{}& L(0)+\int_0^t 
\w(x)[\fr{\sin\tzt\v^2(x)}{2}+\ti R(\v)(x) ]
+\u\dx\v(x)[\fr{\sin\tzt\v^2(x)}{2}+\ti R(\v)(x) ]\\
{}&-\d\u\int \fr{\sin(\tzt)}2\du\tzt\v^2(x)\,dx\\
{}&+ (\d\xi-\u) \int  {\cos(\tzt)} \v(x)\dx\v(x)\,dx+\d\u\int \w(x)\dx\v(x)\,dx\\
{}& 
-\u\d u \g(u)^3\eps^2\int (x-\xi)\t_K'(\g(u)(x-\xi))f(\eps x)\,dx\\ 
{}&+(\u-\d\xi)\g(u)\eps^2\int \t_K'(\g(u)(x-\xi)) f(\eps x)\,dx
-\u\eps^3\int \v(x) f'(\eps x)\,dx
 \,dt\\
 {}&+   \nhone{\v(t )}\eps^{\fr 32}\nltwo{f}+\nhone{\v(0 )} \eps^{\fr 32}\nltwo{f} \\
 \le{}&  L(0)+C\int_0^t 
\eps^{2+ \fr 12+ \fr 5 4}
\,dt + \fr c 8 \nhone{\v(t )}^2 +\fr c 8 \nhone{\v(0 )}^2 + \fr 4 {c}\nltwo{f }^2 \eps^3\,, 
\ee
since
\be
 -\u\eps^3\int \v(x) f'(\eps x)\,dx \le |u|\eps^{ 2+ \fr 12} \nhone{v}\nltwo{f' }\,.
\ee
After bringing two terms on the left hand side we obtain
\be
 {\CR  \ti c}(\tnv{v}{0}{t}^2+\tnw{w}{0}{t}^2) 
\le L(0)+C\int_0^t
\eps^{\fr{15}4}
\,dt  + \fr 4 {c}\nltwo{f}^2 \eps^3 \,.
\ee
\epr
\bth\la{th mainprooftrcontradiction epstwof}
Suppose that the assumptions of \cref{maintheorem epstwof} are satisfied. 
Assume that \re{SGE1 initialdataprob introduction}  has a solution $(\t,\p)$ on $[0,\overline T]$ such that
$$
(\t,\p) \in C^1([0,\overline T], L^\infty(\R) \oplus L^2(\R))\,.
$$
Suppose that
$
0\le T \le \overline T\,.
$
Then, provided 
$
0\le T\le {\eps^{-1 }}, 
$
it holds that $(v,w)$ given by \re{decomposition1 epstwof}-\re{decomposition2 epstwof} is well defined for times $[0,T]$ and 
there exists a constant $\hat c$ such that 
\begin{itemize}
\item[(a)] $\tnv{v}{0}{T}^2+\tnw{w}{0}{T}^2\le \hat c \eps^\fr {11} 4\,,$
\item[(b)] $\forall  t \in [0,T]:~(\xi(t),u(t))\in \Sigma(5,U)\,.$
\end{itemize}
\eth
\bpr
Notice that $\Sigma(5,U)
\subset \Sigma(4,U)$.
We define an exit time
\be
t_*:= \sup\Big\{{}& T>0:\tnv{v}{0}{t}^2+\tnw{w}{0}{t}^2\le 2\bar C ( L(0)+ \eps^\fr {11} 4),\\ 
{}& (\xi(t),\u(t))\in \Sigma(5,U),~0\le t \le T\Big\}\,.
\ee
Suppose $t_* < \fr {1} {\eps^{  }}$. Then there exists a time $\hat t$ such that 
$
\fr {1} {\eps^{ }}>\hat t>t_*\,,
$ 
with
$$ \forall  t \in [0,\hat t]:~(\xi(t),\u(t))\in \Sigma(4,U),~~~~(\xi(\hat t),\u(\hat t))\notin \Sigma(5,U)$$ 
or
$$
\bar C ( L(0)+\eps^{\fr{11}4})<2\bar C ( L(0)+ \eps^{\fr{11}4})<\tnv{v}{0}{\hat t}^2+\tnw{w}{0}{\hat t}^2<\eps^{\fr 5 2}\,.
$$
This leads a contradiction to \cref{le: prepairing main proof epstwof}. 
\epr

\noindent The previous theorem implies that the local solution of \re{SGE1 initialdataprob introduction} discussed in \cref{local solution epstwof} is indeed continuable up to times $  {\eps^{-1 }}$. \cref{th mainprooftrcontradiction epstwof} and \cref{le Modulation Equations epstwof} yield the approximate equations for the parameters $(\xi,\u)$. This concludes the claim of \cref{th: preparing maintheorem epstwof}.
\qed\bigskip
\section{ODE Analysis}\la{se Pt2 ODE analysis}
In this section we lay the groundwork for passing from the approximate equations for the parameters $(\xi,\u)$
in \cref{th: preparing maintheorem epstwof} to the ODEs given by \re{exactODE epstwof1}.
We start with a preparing lemma.
\ble\la{le reference trajectory epstwof}
Let $\tixi=\tixi(s)$, $\tiu=\tiu(s)$, 
$\epsilon_1=\epsilon_1(s)$, $\epsilon_2=\epsilon_2(s)$ be $C^1$ real-valued 
functions.  Suppose that $f\in H^3(\R)$ 
and that 
$$|\epsilon_j| \leq \bar c \eps^{ \fr {3}4}$$ on $[0,T]$ for $j=1,2$. Assume that on $[0,T]$, 
\be
\ds \tixi(s) ={}&  \tiu(s)+\epsilon_1(s)\,,~~\ti\xi(0)=\ti\xi_0\\
\ds \tiu(s) ={}& - \fr {f(\eps\tixi(s))} {[\g(\eps\tiu(s))]^3 m} \int \t_K ' (Z) \,dZ +\epsilon_2(s)\,,~~\ti\u(0)=\ti\u_0.
\ee
Let $\hxi=\hxi(s)$ and $\hu=\hu(s)$ be $C^1$ real-valued 
functions which satisfy the exact equations
\be
\ds \hxi(s) ={}&  \hu(s)  \,,~~\hat\xi(0)=\ti\xi_0\\
\ds \hu(s) ={}& - \fr { f(\eps\hxi(s))} {[\g(\eps\hu(s))]^3 m}\int \t_K ' (Z) \,dZ\,,~~\hat\u(0)=\ti\u_0.  
\ee
Then provided 
$T\leq 1$, there exists $c>0$ such that the estimates
$$|\tixi-\hxi|\leq c\eps^{\fr {3}4 }, 
\qquad |\tiu-\hu| \leq c\eps^{ \fr {3}4} \,.$$
hold on $[0,T]$.

\ele

\bpr
In the following proof we follow very closely \cite[Lemma 6.1]{HoZwSolitonint}.
Let $x=x(t)$ and $y=y(t)$ be $C^1$ 
real-valued functions, $C\ge 1$, and let $(x,y)$ satisfy the differential 
inequalities:
$$
\left\{
\begin{aligned}
&|\dot x| \leq |y| \\
&|\dot y| \leq C |x|+ C |y|
\end{aligned}
\right. ,
\qquad
\begin{aligned}
&x(0)=x_0\\
&y(0)=y_0
\end{aligned}\,.
$$
We are going to apply the Gronwall lemma.
Let $z(t)=x^2+y^2$.  Then 
$$|\dot z| = |2x\dot x + 2y\dot y| \leq 2|x||y| 
+ 2C |x||y| +2C |y||y| \leq 4C(x^2+y^2) = 4Cz\,$$ 
and hence $z(t) \leq z(0)e^{4Ct}$.
Thus
\be  \la{Gronwall epstwof}
 |x(t)| \leq \sqrt 2\max(|x_0|,|y_0|) \exp(2Ct)\,,~~~~
 |y(t)| \leq \sqrt 2\max(|x_0|,|y_0|) \exp(2Ct)\,.
\ee
Now we recall the Duhamel's formula.
Let $X(s): \mathbb{R} \to \mathbb{R}^2$ be a two-vector function,  
$X_0\in \mathbb{R}^2$ a two-vector, and 
$A(s):\mathbb{R}\to (2\times 2\text{ matrices})$ a $2\times 2$ matrix function. We consider the ODE system $$\dot X(s) = A(s)X(s),~~~~X(s')=X_0$$ and denote its solution by $X(s)=S(s,s')X_0$ such that
\[ \frac{d}{ds} S(s,s')X_0 = A(s)S(s,s')X_0 \,, \ \ S(s',s')X_0=X_0 \, . \]
Let $F(s):\mathbb{R}\to \mathbb{R}^2$ be a two-vector function. We can express the 
solution to the inhomogeneous ODE system  
\begin{equation}
\label{E:inhomODE epstwof}
\dot X(s) = A(s)X(s) + F(s)
\end{equation}
with initial condition $X(0)=0$ by the Duhamel's formula
\begin{equation}
\label{E:Duhamel epstwof}
X(s) = \int_0^s S(s,s')F(s')ds'\,.
\end{equation}
Let $U=\hu-\tiu$ and $\Xi=\hxi-\tixi$. These functions satisfy 
\be
\ds \Xi(s) ={}&  U(s) + \epsilon_1(s) \,,\\ 
\ds U(s) ={}& \Big[\fr {f(\eps\hxi(s))} {[\g(\eps\hu(s))]^3 m}-\fr {f(\eps\tixi(s))} {[\g(\eps\tiu(s))]^3 m}\Big]\int \t_K ' (Z) \,dZ + \epsilon_2(s)\,.
\ee
Notice that
\be
{}&\fr {f(\eps\hxi(s))} {[\g(\eps\hu(s))]^3 }-\fr {f(\eps\tixi(s))} {[\g(\eps\tiu(s))]^3 }\\
={}&\fr {f(\eps\hxi(s)) -f(\eps\tixi(s))} {[\g(\eps\hu(s))]^3 }
+\fr{f(\eps\tixi(s))} {[\g(\eps\hu(s))]^3 }
-\fr {f(\eps\tixi(s))} {[\g(\eps\tiu(s))]^3 }\\
={}&\fr 1 {[\g(\eps\hu(s))]^3}\fr {f(\eps\hxi(s)) -f(\eps\tixi(s))}{\hxi(s)-\tixi(s) }
\Big[\hxi(s)-\tixi(s) \Big]\\
{}&+\fr{f(\eps\tixi(s))}{[\g(\eps\tiu(s))]^3[\g(\eps\hu(s))]^3 }\fr {[\g(\eps\tiu(s))]^3-[\g(\eps\hu(s))]^3}
{\tiu(s)-\hu(s)}
\Big[\tiu(s)-\hu(s) \Big] \,.
\ee
Let 
\be
g^\eps(s)={}&\left\{
\begin{aligned}
& \fr 1 {[\g(\eps\hu(s))]^3 m}\int \t_K ' (Z) \,dZ  \fr{{f(\eps\hxi(s))} -{f(\eps\tixi(s))}}{(\hxi(s)-\tixi(s))} & \ \ \text{if }\hxi(s) \neq \tixi(s)\\
& \ \  \fr{\eps f'(\eps\hxi(s))}{[\g(\eps\hu(s))]^3 m}\int \t_K ' (Z) \,dZ  &\ \ \text{if } \hxi(s)=\tixi(s)
\end{aligned}
\right.,
\\
\\
h^\eps(s)={}&\left\{
\begin{aligned}
& \fr{f(\eps\tixi(s))}{[\g(\eps\tiu(s))]^3[\g(\eps\hu(s))]^3m }\int \t_K ' (Z) \,dZ 
\fr {[\g(\eps\tiu(s))]^3-[\g(\eps\hu(s))]^3}
{\tiu(s)-\hu(s)}
 & \ \ \text{if }\hu(s) \neq \tiu(s)\\
& \ \  \fr{3\eps [\g(\eps\hu(s))]^3 \hu(s) f(\eps\tixi(s)) }{[\g(\eps\tiu(s))]^3[\g(\eps\hu(s))]^3m } 
\int \t_K ' (Z) \,dZ 
&\ \ \text{if } \hu(s)=\tiu(s)
\end{aligned}
\right..
\ee
We set
\be
A^\eps(s) =\bmat 0 & 1 \\ 
g^\eps(s) & h^\eps(s) 
\emat, 
\quad F(s) = \bmat \epsilon_1(s) \\ \epsilon_2(s) \emat, 
\quad X(s)=\bmat   \Xi(s) \\ U(s) \emat
\ee
and obtain by Duhamel's formula:
\begin{equation}
\label{E:Duhamel2 epstwof}
\begin{bmatrix}
\Xi(s) \\ U(s)
\end{bmatrix}
= \int_0^s S^\eps(s,t') \begin{bmatrix} \epsilon_1(t') \\ \epsilon_2(t') 
\end{bmatrix} \, dt'.
\end{equation}
Applying \re{Gronwall epstwof} with
$$\begin{bmatrix} x(s) \\ y(s) \end{bmatrix} = S^\eps(s+t',t')\begin{bmatrix} 
\epsilon_1(t') \\ \epsilon_2(t') \end{bmatrix},  \quad \begin{bmatrix} x_0 
\\ y_0 \end{bmatrix} = \begin{bmatrix} \epsilon_1(t') \\ 
\epsilon_2(t') \end{bmatrix}\, $$
yields
$$\left| S^\eps(s,t') \begin{bmatrix} \epsilon_1(t') \\ 
\epsilon_2(t') \end{bmatrix} \right| \leq \sqrt 2 \begin{bmatrix} 
\exp(2C(s-t')) \\ \exp(2C(s-t')) \end{bmatrix} 
\max(|\epsilon_1(t')|,|\epsilon_2(t')|)\,.$$
Using \eqref{E:Duhamel2 epstwof} we obtain that on $[0,T]$
\be
&|\Xi(s)| \leq \sqrt 2 \, T {\exp(2CT)}
\sup_{0\leq t\leq T}\max(|\epsilon_1(t)|,|\epsilon_2(t)|)\,,\\
&|U(s)| \leq \sqrt 2 \, T\exp( 2C T) 
\sup_{0\leq t\leq T}\max( |\epsilon_1(t)|,|\epsilon_2(t)|)\,,
\ee
which proves the claim.
\epr
\noindent
In the following we show the relation between the parameters $(\xi,\u)$ selected by the implicit function theorem according to \cref{le uniform decomposition cl} and the solutions $(\hat\xi,\hat\u)$ of the exact ODEs from the previous lemma.

\ble\la{le exact dynamics epstwof}
Suppose that the assumptions 
of \cref{maintheorem epstwof} are satisfied.
Let $\eps$ be sufficiently small, 
 $s=\eps^{ } t $,   
where 
$$0\le s\le 1,~~0\le t\le \fr 1 {\eps^{}}\,.$$
Let $(\xi,\u)$ be the parameters selected according to \cref{le uniform decomposition cl} and  
$(\hat\xi,\hat\u)$ from \cref{le reference trajectory epstwof}.
Then there exists $c>0$ such that
\be
  |\xi(t)- {\hxi(\eps^{ } t)}|\le c\eps^{ \fr 3 4}\,,~~~~
 |\u(t)-\eps\hu(\eps^{ } t)|  \le c\eps^{\fr 7 4}\,.
\ee
\ele
\bpr
We set
$$\tixi(s)=  \xi(s/\eps^{  }), ~~~~\tiu(s)= \fr 1 \eps u(s/\eps^{  })\,.$$
\noindent
For times
$
0\le t\le  {\eps^{-1}}\,
$
\cref{th: preparing maintheorem epstwof}
yields that
\be
|\d\xi -u |  
\le   C \eps^{\fr {11}4} ,~~~~~~
	|\d\u  + W(\eps,\xi,\u) | 
\le   C \eps^{\fr {11}4} \,. 
\ee
\noindent
Thus $(\tixi,\tiu)$ satisfy the assumptions of \cref{le reference trajectory epstwof}, 
which implies that
\be
| \xi(t)-\hxi(\eps^{ } t)|=   |\tixi(s)-\hxi(s)|\le c\eps^{ \fr {3}4},~~~~~~
 |\fr 1 \eps\u(t)-\hu(\eps t)|=  |\tiu(s)-\hu(s)|  \le c \eps^{ \fr {3}4 }.
\ee 
\epr
\section{Completion of the Proof of \cref{maintheorem epstwof}}\la{Completion of the Proof of Theorem}
\cref{th: preparing maintheorem epstwof} yields the dynamics with the parameters $(\xi,\u)$  selected by the implicit function theorem according to \cref{le uniform decomposition cl} on the time interval 
$
0\le t\le {\eps^{-1  }}\,.
$
Using \cref{le exact dynamics epstwof} and the the triangle inequality we can replace $(\xi(t),\u(t))$
with $(\bar\xi(t),\bar\u(t)):=( {\hxi(\eps^{ } t)},\eps^{ } \hu(\eps^{ } t))\,.$
We have to replace $\eps^{\fr {11} 4}$ with $\eps^{ \fr {3}2}$ in the upper bound on the squared norm of the transversal component $(\v,\w)$
since the difference of the parameters $|\xi(t)- {\hxi(\eps^{ } t)} |$ in \cref{le exact dynamics epstwof} is of order $\eps^{ \fr {3}4}$.
\qed\bigskip

\end{document}